\documentclass[12pt]{iopart}

\usepackage{bm,graphicx,amsfonts}

\def\be{\begin{equation}}
\def\ee{\end{equation}}
\def\v{\bm}
\def\vc{{\v c}}
\def\vv{{\v v}}
\def\vx{{\v x}}
\def\vf{{\v f}}
\def\vA{{\v A}}
\def\vell{{\v \ell}}
\def\vr{{\v r}}
\def\sgn{{\mathop{\rm sgn}}}
\def\cross{{\v\times}}
\def\vc{{\v c}}
\def\vrho{{\v \rho}}
\def\vn{{\v n}}
\def\vC{{\v C}}
\def\vp{{\v p}}
\def\R{\mathbb R}
\def\goesto{\rightarrow}

\def\Eq#1{(\ref{#1})}
\def\Eqs#1{(\ref{#1})}

\def\hEq#1{~(\ref{#1})}

\def\x{\xi}
\def\t{\tau}
\def\fr#1#2{\frac{#1}{#2}}
\def\O{{\cal O}}
\def\inv{^{-1}}

\def\({\left(}\def\){\right)}
\def\[{\left[}\def\]{\right]}

\newtheorem{lemma}{Lemma}

\begin{document}

\title{An Exactly Conservative Integrator for the $n$-Body Problem}

\author{Oksana Kotovych and John C. Bowman}

\address{Department of Mathematical and Statistical Sciences,
University of Alberta, Edmonton, Alberta, Canada T6G~2G1}

\ead{bowman@math.ualberta.ca}

\date{\today}

\begin{abstract}  
The two-dimensional $n$-body problem of classical mechanics is a
non-integrable Hamiltonian system for $n > 2$. Traditional numerical
integration algorithms, which are polynomials in the time step, typically lead
to systematic drifts in the computed value of the total energy and
angular momentum. Even symplectic integration schemes exactly conserve only
an approximate Hamiltonian. We present an algorithm that conserves the true
Hamiltonian and the total angular momentum to machine precision.
It is derived by applying conventional discretizations in a new space
obtained by transformation of the dependent variables. We develop the
method first for the restricted circular three-body problem, then for the
general two-dimensional three-body problem, and finally for the planar
$n$-body problem. Jacobi coordinates are used to reduce the two-dimensional
$n$-body problem to an $(n-1)$-body problem that incorporates the constant
linear momentum and center of mass constraints. For a four-body choreography,
we find that a larger time step can be used with our conservative algorithm
than with symplectic and conventional integrators.
\end{abstract}

\submitto{\JPA\rm 22 Dec 2001; accepted in revised form 02 Aug 2002}



\maketitle

\section{Introduction}

The $n$-body problem is the study of the motion of $n$ arbitrary particles in
space according to the Newtonian law of gravitation.  When $n=2$ (the
Kepler problem), the problem has a well-known analytic solution, but
Poincar\'e has shown that the system is in general non-integrable for $n > 2$.
To approximately solve these cases, one often attempts to discretize the
equations of motion and study the evolution of the system
numerically. However, discretization of a system of differential
equations typically leads to a loss of accuracy; first integrals of the
motion may no longer be preserved and the phase portrait may become inaccurate.
This often necessitates the use of small time steps, so that many iterations
will be required. In this article, we
demonstrate that {\em conservative integration} can be used to obtain an
accurate picture of the dynamics even with a relatively large time step.

Conservative integration was introduced by Shadwick, Bowman, and Morrison
\cite{Shadwick99,Bowman97,Shadwick01}. These authors argued
that a more robust and faithful evolution of the dynamics can be obtained
by explicitly building in knowledge of the analytical structure of the
equations; in this case, by preserving the known first integrals of the
motion. They illustrated the method applied to a three-wave truncation of
the Euler equations, the Lotka--Volterra problem, and the Kepler
problem. In this work, we extend the method to the equations of motion of
$n$ bodies in space, first to the circular restricted three-body problem,
then to the general three-body problem, and finally to the full $n$-body
case.  For simplicity we only consider two-dimensional motion (a reasonable
assumption for all of the planets in the solar system except for Pluto);
extending this work to three dimensions should be straightforward.

\section{Conservative Integration}
The equations describing the motion of the solar system form a conservative
system: the friction that heavenly bodies sustain is so small that
virtually no energy is lost. Both the total energy and total angular
momentum are conserved.  We argue that a robust integration algorithm
should preserve both of these invariants.  

One way to accomplish this is to transform the dependent variables to a new
space where the energy and other conserved quantities are linear functions
of the transformed variables, apply a traditional integration algorithm in this
space, and then transform back to get new values for each variable
\cite{Shadwick99,Bowman97}. This approach is motivated by the following
trivial lemma.

\begin{lemma}\label{linear}
Let $\vx$ and $\vc$ be vectors in $\R^n$. If $\vf:\R^{n+1}\goesto\R^n$
has values orthogonal to $\vc$, so that $I=\vc\cdot\vx$ is a linear
invariant of the first-order differential equation $d\vx/dt=\vf(\vx,t)$, then
each stage of the explicit $m$-stage discretization
\be
\vx_j=\vx_0+\tau \sum_{k=0}^{j-1} b_{jk}\vf(\vx_k,t+a_j\tau),\qquad j=1,\ldots, m,
\label{m-stage}
\ee
also conserves $I$, where $\tau$ is the time step and $b_{jk}\in \R$.
\end{lemma}
Proof.
For $j=1,\ldots,m$, we have $\vc\cdot\vx_j=\vc\cdot \vx_0+\tau
\sum_{k=0}^{j-1} b_{jk}\vc\cdot 
\vf(\vx_k,t+a_j\tau)=\vc\cdot \vx_0.\quad\diamond$

A conservative integration algorithm can be constructed by writing
any conventional integration algorithm of the form\hEq{m-stage}, for which
specific values of $a_j$ and $b_{jk}$ are known, in a transformed space.
For example, 
consider the \emph{second-order predictor--corrector} (2-stage) scheme
for evolving the system of ordinary differential equations
$d\vx/dt=\v{f}(\vx,t)$,
\numparts
\be\label{predictorcorrector}
\tilde{\vx}=\vx_{0}+\tau \v{f}(\vx_0,t), \label{predictorcorrector a}
\ee
\be
\v{x}(t+\tau)=\v{x}_0+\frac{\tau}{2} [ \v{f}(\vx_0,t)+ \v{f}(\tilde{\vx},t+\tau)], \label{predictorcorrector b}
\ee
\endnumparts
where we now write $\tilde \vx$ instead of $\vx_1$.
In the \emph{conservative predictor--corrector algorithm},
one seeks a transformation $\v{\xi}=\v{T}(\v{x})$ of the dependent variable
$\vx$ such that the quantities to be conserved can be expressed as linear
functions of the new variables $\xi_{i}$, $i=1,\ldots ,n.$  Then, keeping
\Eq{predictorcorrector a} as the predictor, in the transformed space one
applies the corrector 
\be
\v{\xi}(t+\tau)=\v{\xi}_0+\frac{\tau}{2} [\v{T}'(\v{x_0}) \v{f}(\vx_0,t)+
\v{T}'(\v{\tilde{x}}) \v{f}(\tilde{\vx},t+\tau)], \label{cpcb}
\ee
where $\v \xi_0=\v T(\vx_0)$ and $\v{T}'$ is the derivative of $\v{T}$.
The new value of $\vx$ is obtained by inverse transformation,
$\v{x}(t+\tau)=\v{T}^{-1}(\v{\xi}(t+\tau))$.
Often the inverse transformation involves radicals, and if the argument of the
radical becomes negative, it is possible to use a finite number of
time-step reductions to integrate the system through this region
\cite{Bowman97}; this approach is particularly advantageous when the time
step is chosen adaptively. Another way to deal with noninvertible
transformations is to switch to a conventional (e.g.\ predictor--corrector)
integrator for that one time step. 
If the inverse transformation involves several branches (e.g.\ because of a
square root), the correct branch can be distinguished with sufficient accuracy
using the conventional predictor solution. The error analysis for a
second-order predictor--corrector algorithm is described
in~\ref{error}. Higher-order conservative integration algorithms are
readily obtained in the same way, by coupling the first $m-1$ ``predictor''
stages from \Eq{m-stage} with the final conservative corrector stage
\be
\v{\xi}(t+\tau)=\v{\xi}_0+\tau
\sum_{k=0}^{m-1} b_{mk}\v{T}'(\v{x_k}) \v{f}(\vx_k,t+a_j\tau).
\ee

According to Iserles \cite{Iserles97}, a major drawback of traditional
non-conservative integration is that numbers are often ``thrown into the
computer.'' Mathematical models are often discretized according to algorithms
that have little to do with the original problem. Iserles argued that one
should develop computational algorithms that reflect known structural features
of the problem under consideration (e.g.\ see \cite{deFrutos94,Shadwick00}). The
conservative predictor--corrector is an example of such an integrator.  In
the examples given by \cite{Shadwick99,Shadwick01}, the transformation
$\v{T}$ is tailored to the system at hand; there is obviously no generic
transformation that can be used to integrate an arbitrary conservative system.

It is interesting to compare conservative integration (which conserves the
value of the Hamiltonian) with symplectic integration (which conserves
phase-space volume; see Refs.~\cite{Ruth83}, \cite{Forest90},
\cite{Cooper87}, \cite{Channell-Scovel90}, and~\cite{Sanz-Serna-Calvo94}).
According to Ge and Marsden (1988)\nocite{Ge88}, if an integrator is both
symplectic and conservative, it must be exact. Normally we do not have the
luxury of an exact discretization at our disposal. The drawback then with
conservative integration is that the Hamiltonian phase-space structure
will not be preserved, just as for symplectic integration the
total energy will not be conserved. Which method is preferable depends on the
physical structure of the problem being investigated. 

Another important advantage of conservative integration algorithms is that,
unlike typical symplectic integration schemes, they are explicit. Although in
some cases the inverse of the transformation $T$ may be defined by
an implicit equation that requires iteration to solve (using the predicted
value as an accurate initial guess), this is really nothing more than a
special function evaluation; the time-stepping scheme itself, being
causal, is explicit.

With conservative integration, one can preserve all of the known invariants
of the $n$-body problem conserved exactly, even for large time steps. This
can lead to a more accurate picture of the motion of the bodies
\cite[figure 9]{Shadwick99} for the same computational effort. In the next
section, we motivate the extension of the method of conservative
integration to the $n$-body problem by briefly revisiting the treatment of
the Kepler problem in Ref.~\cite{Shadwick99}. 

\section{Kepler Problem}
The Kepler problem describes the motion of two bodies with masses $m_1$
and $m_2$ located at positions $\vr_1$ and $\vr_2$, respectively.
The dynamics can be reduced to an equivalent one-body problem, the
behaviour of a single particle of mass $m=m_1m_2/(m_1+m_2)$ at the position
$\vr = \vr_2 - \vr_1$ under the influence of a central gravitational
force. This force may be expressed as the gradient of the potential
function $V=-k/r$, where $k=Gm_1m_2$ and~$G$ is the universal gravitational
constant. The equations of motion can be written in terms of the radial
velocity~$v_r$ and the polar coordinate angle~$\theta$ of the particle,
\numparts
\be\label{drdvdthdl}
\frac{dr}{dt}=v_{r}, \label{drdvdthdl a}
\ee
\be
\frac{dv_{r}}{dt}=\frac{\ell^{2}}{m^2r^{3}}-\frac{1}{m}\(\frac{\partial V}{\partial r}\), \label{drdvdthdl b}
\ee
\be
\frac{d\theta}{dt}=\frac{\ell}{mr^{2}}, \label{drdvdthdl c}
\ee
\endnumparts
where $\ell$ is the (constant) total angular momentum. It is convenient to
rewrite the equations in terms of the linear momentum $p=mv_{r}$ and the
angular momentum~$\ell$:
\numparts
\be
\frac{dr}{dt}=\frac{\partial H}{\partial p}=\frac{p}{m},
\ee
\be
\frac{dp}{dt}=-\frac{\partial H}{\partial r}=\frac{\ell^{2}}{mr^{3}}-\frac{\partial V}{\partial r},
\ee
\be
\frac{d\theta}{dt}=\frac{\partial H}{\partial \ell}=\frac{\ell}{mr^{2}},
\ee
\be
\frac{d\ell}{dt}=-\frac{\partial H}{\partial \theta}=0,
\ee
\endnumparts
where the Hamiltonian
\be
H=\frac{p^{2}}{2m}+\frac{\ell^{2}}{2mr^{2}}+V(r), \label{Hamiltonian}
\ee
is also conserved.

\subsection{Integration}
To set the framework for extending two-body conservative integrators to
the $n$-body problem, we slightly generalize the presentation in
Ref.~\cite{Shadwick99} to make the constant~$\ell$ a variable that is
formally integrated, but which remains constant.

The predictor step of the conservative integrator is given
by~\Eq{predictorcorrector a}, where $\vx=(r,\theta,p,\ell)$. To derive
the corrector, the vector $(r,p,\ell)$ is transformed to
$(\xi_1,\xi_2,\xi_3)$, where
\numparts
\be\label{Keplertransformedequations}
\xi_{1}=-\frac{k}{r}, \label{Keplertransformedequations a}
\ee
\be
\xi_{2}=\frac{p^{2}}{2m}+\frac{\ell^{2}}{2mr^{2}}, \label{Keplertransformedequations b}
\ee
\be
\xi_{3}=\ell. \label{Keplertransformedequations c}
\ee
\endnumparts
On differentiating these equations with respect to time and exploiting the
fact that both $H=\xi_{1}+\xi_{2}$ and $L=\xi_{3}$ are conserved, one
finds
\numparts
\be\label{Keplerderivativesoftransformedequations}
\dot{\xi}_{1}=\frac{kp}{mr^{2}},  \label{Keplerderivativesoftransformedequations a}
\ee
\be
\dot{\xi}_{2}=-\dot{\xi}_{1},  \label{Keplerderivativesoftransformedequations b}
\ee
\be
\dot{\xi}_{3}=0.  \label{Keplerderivativesoftransformedequations c}
\ee
\endnumparts
After applying \Eq{cpcb}, the inverse transformation
\numparts
\be\label{Keplerinversetransformation}
r=-\frac{k}{\xi_{1}}, \label{Keplerinversetransformation a}
\ee
\be
\ell=\xi_{3}, \label{Keplerinversetransformation b}
\ee
\be
p=\sgn(\tilde{p})\sqrt{2m\xi_{2}-\frac{\ell^{2}}{r^{2}}} \label{Keplerinversetransformation c}
\ee
\endnumparts
is used to update the values of the original variables at the new step.
See Ref.~\cite{Shadwick99} for details on how the invariance of the
Runge--Lenz vector $\vA=\vv \v{\times} \vell + V\vr$ is exploited to
evolve~$\theta$.\footnote{There is a typographical error in Eq. (54b) of
Ref.~\cite{Shadwick99}; it should read
\be
v_r(t+\tau) =\sgn(\widetilde{v_r})\,
\sqrt{v_r^2+\frac{\ell^2}{m^2}\left(\frac{1}{r^2} -
\frac{1}{r^2(t+\tau)}\right)-2\,\frac{\Delta}{m}}.
\ee
}

Before generalizing the integrator of Shadwick {\it et al.\/} to the $n$-body
problem, it is instructive to consider first the special case of the
restricted three-body problem.

\section{Restricted Three-Body Problem} 

Suppose that two bodies of masses $m_{1}$ and $m_{2}$, called the
primaries, revolve around their center of mass in circular orbits. 
The \emph{circular restricted three-body problem} describes the motion of
a third body, with a mass $m_{3}$ that is negligible
compared to $m_{1}$ and $m_{2}$, at coordinates $(x,y)$ in the plane of motion
of the other two bodies. 
The third body does not influence the motion of the other two. The
derivation of the equations of motion for the restricted problem is
described in \cite{Szebehely67}. The Hamiltonian is given by
\be
H=\frac{1}{2} \(\dot{x}^{2}+\dot{y}^{2}\)-\frac{1}{2} \(y^{2}+x^{2}\)-\frac{1-\mu}{r_{1}}-\frac{\mu}{r_{2}},
\ee
where $r_{1}^2=(x-\mu)^{2}+y^{2}$, $r_{2}^2=(x+1-\mu)^{2}+y^{2}$, and
$\mu=m_2/(m_1+m_2)$.
In terms of the canonical variables
\be
q_{1}=x,\qquad q_{2}=y,\qquad p_{1}=\dot{x}-y,\qquad p_{2}=\dot{y}+x,
\ee
the Hamiltonian appears as
\be
H=\frac{1}{2} (p_{1}^{2}+p_{2}^{2})+p_{1}q_{2}-p_{2}q_{1}-\frac{1-\mu}{r_{1}}-\frac{\mu}{r_{2}}.
\ee
The equations of motion are then
\numparts
\be\label{r3beqmot}
\dot{q}_{1}=\frac{\partial H}{\partial p_{1}}=p_{1}+q_{2},\label{r3beqmot a}
\ee
\be
\dot{q}_{2}=\frac{\partial H}{\partial p_{2}}=p_{2}-q_{1},\label{r3beqmot b}
\ee
\be
\dot{p}_{1}=-\frac{\partial H}{\partial q_{1}}=p_{2}-\frac{1-\mu}{r_{1}^3}(q_{1}-\mu)-\frac{\mu}{r_{2}^3}(q_{1}+1-\mu),\label{r3beqmot c}
\ee
\be
\dot{p}_{2}=-\frac{\partial H}{\partial q_{2}}=-p_{1}-\frac{1-\mu}{r_{1}^3}q_{2}-\frac{\mu}{r_{2}^3}q_{2},\label{r3beqmot d}
\ee
\endnumparts
and the Hamiltonian can be rewritten as
\be
H=\frac{1}{2} \(\dot{q}_{1}^{2}+\dot{q}_{2}^{2}\)-\frac{1}{2} \(q_{1}^{2}+q_{2}^{2}\)-\frac{1-\mu}{r_{1}}-\frac{\mu}{r_{2}}. \label{r3bHamiltonian}
\ee

\subsection{Integration}
The conventional predictor for this system is
\be
\tilde{q}_{i}=q_{i}+\dot{q}_{i}\tau,
\qquad \tilde{p}_{i}=p_{i}+\dot{p}_{i}\tau,
\ee
for $i=1,2$.
Note that, unless specified otherwise, the variables are functions of $t$.
Let 
\numparts
\be\label{r3btransformations}
\xi_{1}=\frac{1}{2}q_{1}^{2},\label{r3btransformations a}
\ee
\be
\xi_{2}=\frac{1}{2}q_{2}^{2},\label{r3btransformations b}
\ee
\be
\xi_{3}=\frac{1}{2}\dot{q}_{1}^{2}-\frac{1-\mu}{r_{1}}-\frac{\mu}{r_{2}},\label{r3btransformations c}
\ee
\be
\xi_{4}=\frac{1}{2}\dot{q}_{2}^{2}.\label{r3btransformations d}
\ee
\endnumparts
Here
\be
H=-\xi_{1}-\xi_{2}+\xi_{3}+\xi_{4} \label{r3blinearH}
\ee
is a linear function of the $\xi$s.
Differentiating the $\xi$s with respect to time, we get
\numparts
\be\label{r3bderivativetransformation}
\dot{\xi}_{1}=q_{1}\dot{q}_{1}, \label{r3bderivativetransformation a}
\ee
\be
\dot{\xi}_{2}=q_{2}\dot{q}_{2}, \label{r3bderivativetransformation b}
\ee
\be
\dot{\xi}_{4}=\dot{q}_{2}\ddot{q}_{2}=\dot{q}_{2}(\dot{p}_{2}-\dot{q}_{1}), \label{r3bderivativetransformation c}
\ee
\be
\dot{\xi}_{3}=\dot{\xi}_{1}+\dot{\xi}_{2}-\dot{\xi}_{4}, \label{r3bderivativetransformation d}
\ee
\endnumparts
on making use of \Eq{r3blinearH}, together with the conservation of $H$.
The corrector is given by
\be
\xi_{i}(t+\tau)=\xi_{i}+\frac{\tau}{2}(\dot{\xi}_{i}+\dot{\tilde{\xi}}_{i}), \label{correctorforr3bbeforethetransformationtakesplace}
\ee
for $i=1,\ldots,4$, where $\tilde{\xi}_{i}$ is simply
\Eq{r3btransformations} evaluated at $\tilde{q}_{i}$, $\tilde{p}_{i}$ and
$t+\tau$. Inverting, the new values of $q_{i}$ and $p_{i}$ can be expressed in
terms of $\xi_{i}$ as
\numparts
\be\label{r3binvertedfirst}
q_{1}=\sgn(\tilde{q}_{1})\sqrt{2\xi_{1}}, \label{r3binvertedfirst a}
\ee
\be
q_{2}=\sgn(\tilde{q}_{2})\sqrt{2\xi_{2}}, \label{r3binvertedfirst b}
\ee
\endnumparts
and, on using \Eqs{r3beqmot a} and\hEq{r3beqmot b},
\numparts
\be\label{r3binvertedsecond}
p_{1}=-q_{2}+\sgn(\tilde{p}_{1}+\tilde{q}_{2})\sqrt{2\xi_{3}+\frac{2(1-\mu)}{r_{1}}+\frac{2\mu}{r_{2}}}, \label{r3binvertedsecond a}
\ee
\be
p_{2}=q_{1}+\sgn(\tilde{p}_{2}-\tilde{q}_{1})\sqrt{2\xi_{4}}. \label{r3binvertedsecond b}
\ee
\endnumparts


This example assumes that the mass of one body is negligible to the 
other two masses and that the other two masses are travelling in circular
orbits. The rest of this paper discusses the general case of three or more 
bodies: no restrictions are placed on the masses of the bodies, and their
orbits do not have to be circular, or even periodic.


\section{General Three-Body Problem}
The derivation of the equations of motion of the general three-body problem
in a plane is described in Refs.~\cite{Szebehely67}, \cite{Barrow-Green97}, 
and \cite{Kovalevsky67}.

Given three bodies $m_{1}$, $m_{2}$, and $m_{3}$ with position vectors
$\vr_1,\ $ $\vr_2,\ $ and  $\vr_3,\ $
where each $\vr_i$ is at location $(x_{i},y_{i}),$ 
define $\vr_{ij} = \vr_j - \vr_i,$
for $i,j = 1,2,3$.
The potential is
\be
V = -\frac{Gm_{1}m_{2}}{r_{12}} - \frac{Gm_{2}m_{3}}{r_{23}} - \frac{Gm_{1}m_{3}}{r_{13}}, \label{g3bV}
\ee
where $G$ is the gravitational constant and $r_{ij} = \sqrt{(x_{j} -
x_{i})^2 + (y_{j} - y_{i})^2}$ is the distance between the $i$th and $j$th bodies.

The system consists of three second-order differential equations,
\numparts
\be\label{equationsofmotioninexpandedform}
m_{1}\ddot{\vr}_1 = -\frac{\partial V}{\partial \vr_1}=\frac{Gm_{1}m_{2}(\vr_2-\vr_1)}{r_{12}^3} +  \frac{Gm_{1}m_{3}(\vr_3-\vr_1)}{r_{13}^3}, \label{equationsofmotioninexpandedform a}
\ee
\be
m_{2}\ddot{\vr}_2 =  -\frac{\partial V}{\partial \vr_2}=\frac{Gm_{1}m_{2}(\vr_1-\vr_2)}{r_{21}^3} +  \frac{Gm_{2}m_{3}(\vr_3-\vr_2)}{r_{23}^3}, \label{equationsofmotioninexpandedform b}
\ee
\be
m_{3}\ddot{\vr}_3 =  -\frac{\partial V}{\partial \vr_3}=\frac{Gm_{1}m_{3}(\vr_1-\vr_3)}{r_{31}^3} +  \frac{Gm_{2}m_{3}(\vr_2-\vr_3)}{r_{32}^3}. \label{equationsofmotioninexpandedform c}
\ee
\endnumparts
These equations conserve the total linear momentum
$\sum_{i=1}^3 m_{i}\dot{\vr}_i$ (which allows us to fix the center of mass
at the origin) and total angular momentum
$\sum_{i=1}^3 \vr_i\cross m_{i}\dot{\vr}_i$. The Hamiltonian
\be
H=\sum_{i=1}^3 m_{i}\dot{\vr}_i^{2} + V, \label{g3bH}
\ee
where $V$ is given by~\Eq{g3bV}, is also conserved. We exploit the
constancy of the linear momentum and center of mass position to reduce the
number of degrees of freedom in the problem. It is convenient to implement
this reduction by converting to Jacobi coordinates (e.g., see
Refs.~\cite{Khilmi61},
\cite{Pollard66}, and~\cite{Roy88}). The remaining constraints of constant
total angular momentum and energy are built into the conservative
integrator by transforming to a frame where these invariants are linear.

Letting $\vr = \vr_2-\vr_1=(r_{x},r_{y})$, $M = m_{1}+m_{2}+m_{3}$, and
$\mu=m_{1}+m_{2}$, the location of the center of mass of $m_{1}$ and
$m_{2}$ is seen to be at $\mu^{-1}(m_{1}\vr_1+m_{2}\vr_2)$, or, since
$m_{1}\vr_1+m_{2}\vr_2+m_{3}\vr_3=\v0$, at
 $-\mu^{-1}m_{3}\vr_3$.
Let $\vrho=(\rho_{x},\rho_{y})$ be the vector from the center of mass of
the first two bodies to the third body. Then 
$\vrho=\vr_3+\mu^{-1}m_{3}\vr_3=M\mu^{-1}\vr_3$ and we find
\numparts
\be\label{whateverthisis}
\vr_2-\vr_1=\vr,\label{whateverthisis a}
\ee
\be
\vr_3-\vr_1=\vrho+m_{2}\mu^{-1}\vr,\label{whateverthisis b}
\ee  
\be
\vr_3-\vr_2=\vrho-m_{1}\mu^{-1}\vr.\label{whateverthisis c}
\ee
\endnumparts
In these coordinates, following \Eq{g3bH}, the Hamiltonian can be written as
\be
H=\frac{1}{2}g_1(\dot{r}_x^2+\dot{r}_y^2) + \frac{1}{2}g_2(\dot{\rho}_x^2+\dot{\rho}_y^2) + V \label{runningoutofnames}
\ee
in terms of the reduced masses $g_{1}=m_{1}m_{2}\mu^{-1}$ and 
$g_{2}=m_{3}\mu/M$, where $V$ is given by~\Eq{g3bV}. 

Define $r_{x}=r\cos\theta$, $r_{y}=r\sin\theta$, $\rho_{x}=\rho\cos\Theta$,
and $\rho_{y}=\rho\sin\Theta$.
In these polar coordinates, the Hamiltonian can be rewritten
\be
H=\frac{p^{2}}{2g_{1}}+\frac{P^{2}}{2g_{2}}+\frac{\ell^{2}}{2g_{1}r^{2}}+\frac{L^{2}}{2g_{2}\rho^{2}}+V(r,\rho,\theta,\Theta),\label{g3bHrewritten}
\ee
where $p$ is the linear momentum of the first reduced mass, $\ell$ is the angular 
momentum of the first reduced mass,
$P$ is the linear momentum of the second reduced mass, $L$ is the angular momentum 
of the second reduced mass, 
and $V=V(r,\rho,\theta,\Theta)$ is the potential energy of the system.
The Hamiltonian $H$ and the total angular momentum $\ell+L$ are
conserved, and the center of mass remains at the origin for all time.

 The equations of motion in polar coordinates are
\numparts
\be
\dot{r}=\frac{\partial H}{\partial p}=\frac{p}{g_{1}},
\qquad
\dot{\theta}=\frac{\partial H}{\partial \ell}=\frac{\ell}{g_{1}r^{2}},
\ee
\be
\dot{p}=-\frac{\partial H}{\partial r}=\frac{\ell^{2}}{g_{1}r^{3}}-\frac{\partial V}{\partial r},
\qquad
\dot{\ell}=-\frac{\partial H}{\partial \theta}=-\frac{\partial V}{\partial \theta},
\ee
\be
\dot{\rho}=\frac{\partial H}{\partial P}=\frac{P}{g_{2}},
\qquad
\dot{\Theta}=\frac{\partial H}{\partial L}=\frac{L}{g_{2}\rho^{2}},
\ee
\be
\dot{P}=-\frac{\partial H}{\partial \rho}=\frac{L^{2}}{g_{2}\rho^{3}}-\frac{\partial V}{\partial \rho},
\qquad
\dot{L}=-\frac{\partial H}{\partial \Theta}=-\frac{\partial V}{\partial \Theta}.
\ee
\endnumparts

\subsection{Integration}

The variables can be transformed as
\numparts
\be
\xi_{1}=\frac{p^{2}}{2g_{1}}+\frac{\ell^{2}}{2g_{1}r^{2}},
\qquad
\xi_{2}=\frac{P^{2}}{2g_{2}}+\frac{L^{2}}{2g_{2}\rho^{2}},
\ee
\be
\xi_{3}=V,\qquad \xi_{4}=\rho,\qquad \xi_{5}=\ell,\qquad \xi_{6}=L,
\qquad \xi_{7}=\theta,\qquad \xi_{8}=\Theta,
\ee
\endnumparts
so that the conserved Hamiltonian becomes a linear function of the
transformed variables: $H=\xi_{1}+\xi_{2}+\xi_{3}$.
The time derivatives become
\numparts
\be
\dot{\xi}_{1}=\frac{p\dot{p}}{g_{1}} + \frac{\ell r^{2} \dot{\ell} - r \ell^{2} \dot{r}}{g_{1}r^{4}},
\ee
\be
\dot{\xi}_{2}=\frac{P\dot{P}}{g_{2}} + \frac{L \rho^{2} \dot{L} - \rho L^{2} \dot{\rho}}{g_{2}\rho^{4}},
\ee
\be
\dot{\xi}_{3}=\frac{\partial V}{\partial r}\dot r+
\frac{\partial V}{\partial \theta}\dot \theta+
\frac{\partial V}{\partial \rho}\dot \rho+
\frac{\partial V}{\partial \Theta}\dot \Theta,
\ee
\be
\dot{\xi}_{4}=\dot{\rho},
\qquad
\dot{\xi}_{5}=\dot{\ell},
\qquad
\dot{\xi}_{6}=\dot{L},
\qquad
\dot{\xi}_{7}=\dot{\theta},
\qquad
\dot{\xi}_{8}=\dot{\Theta}.
\ee
\endnumparts
The integration procedure is an extension of the method used for the
Kepler problem. We can invert to find the original variables as follows,
\numparts
\be
\rho={\xi}_{4},
\qquad
\ell=\xi_{5},
\qquad
L=\xi_{6},
\qquad
\theta=\xi_{7},
\qquad
\Theta=\xi_{8},
\ee
\be
r=g(\xi_3,\rho,\theta,\Theta),
\ee
\be
p=\sgn(\tilde{p})\sqrt{2g_{1}\(\xi_{1}-\frac{\ell^{2}}{2g_{1}r^{2}}\)},
\ee
\be
P=\sgn(\tilde{P})\sqrt{2g_{2}\(\xi_{2}-\frac{L^{2}}{2g_{2}\rho^{2}}\)}.
\ee
\endnumparts
The value of the inverse function $g$ defined by
$V(g(\xi_3,\rho,\theta,\Theta),\rho,\theta,\Theta)=\xi_3$ is determined
at fixed~$\rho$,~$\theta$,~$\Theta$ by Newton--Raphson iteration,
using the predicted value $\tilde r$ as an initial guess.

\begin{figure}
\begin{center}
\includegraphics[height=7.5cm]{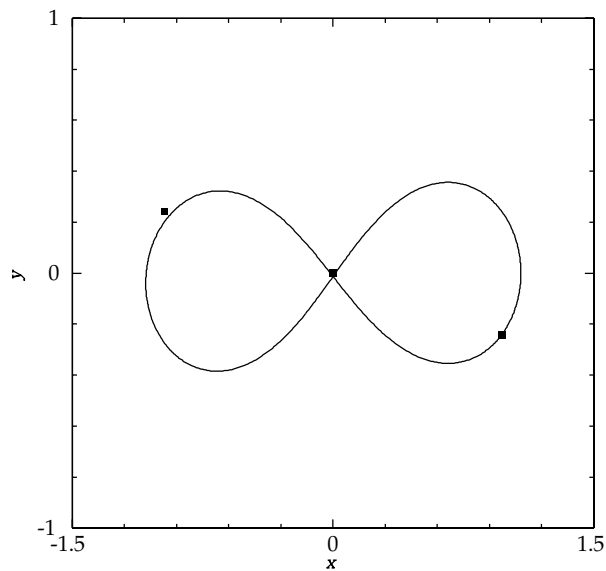}
\caption{Conservative predictor--corrector solution for a general
three-body choreography.}
\label{CPCgeneralthreebody}
\end{center}
\end{figure}

In Fig.~\ref{CPCgeneralthreebody} we use our conservative predictor--corrector
to illustrate the remarkable three-body figure-eight {\it choreography\/} 
discovered by Chenciner and R. Montgomery \cite{Chenciner00} and
located numerically by Sim\'{o} \cite{Simo00}. The dots indicate the
initial positions of the three unit masses. The gravitational constant
$G$ is taken to be unity and the initial conditions are those cited in
\cite{Chenciner00}:
\begin{eqnarray}
\v r_1=(0.97000436,-0.24308753),\quad
\v r_2=(0,0),\nonumber\\
\v r_3=(-0.97000436,0.24308753).\nonumber\\
\dot{\v r_1}=(0.46620369,0.43236573),\quad
\dot{\v r_2}=(-0.93240737,-0.86473146),\nonumber\\
\dot{\v r_3}=(0.46620369,0.43236573).
\end{eqnarray}
We used a fixed time step of $\tau=10^{-4}$ and integrated for a complete
choreographic period, during which each mass travels once around the
figure eight. 

\section{General $\vn$-Body Problem} 

The Jacobi coordinates can be extended to $n\ge 2$ bodies in a plane
\cite{Roy88, Khilmi61}.
Let each body of mass $m_{i}$ have radius vectors
$\vr_i$, where $i=1,\ldots,n$. Define
$\vr_{ij}=\vr_j-\vr_i$ as the vector joining $m_{i}$ to
$m_{j}$. Also define $\vC_i$ to be the center of mass of the
first~$i$ bodies, where $i=2,\ldots,n$, and choose the origin
of the coordinate system so that  $\vC_n=\v0.$ 
Let the vectors $\v{\rho}_i$ be defined such that 
\numparts
\be\label{hownbodyequationsevolve}
\v{\rho}_2=\vr_{12}, \label{hownbodyequationsevolve a}
\ee
\be
\v{\rho}_3=\vr_{3}-\vC_{2}, \label{hownbodyequationsevolve b}
\ee
\be
\ldots
\ee
\be
\v{\rho}_{n}=\vr_n-\vC_{n-1}. \label{hownbodyequationsevolve c}
\ee
\endnumparts
Also,
\be
\vr_{k\ell}=\v{\rho}_\ell-\v{\rho}_k+\sum_{j=k}^{\ell-1} \frac{m_{j}\v{\rho}_{j}}{M_{j}}, \label{generalformoftheaboveequatiosforthenbodyproblem}
\ee
where $1\le k < \ell \le n,$ and 
$M_{j}=\sum_{k=1}^{j-1} m_{k}$.\footnote{Here $\v{\rho}_1$ 
is a dummy variable that  cancels out in the expression for $r_{12}.$}

The reduced masses are 
\be
g_{2}=\frac{m_{2}m_{1}}{M_{2}},
\ee
\be
g_{3}=\frac{m_{3}(m_{2}+m_{1})}{M_{3}},
\ee
\be
\ldots
\ee
\be
g_{n}=\frac{m_{n}M_{n-1}}{M_{n}}.
\ee

The equations of motion in polar coordinates are just an extension of the three-body problem:
\numparts
\be\label{gnbrthpl}
\dot{\rho}_{i}=\frac{\partial H}{\partial p_{i}}=\frac{p_{i}}{g_{i}},\label{gnbrthpl a}
\ee
\be
\dot{\theta}_{i}=\frac{\partial H}{\partial \ell_{i}}=\frac{\ell_{i}}{g_{i}\rho_{i}^{2}},\label{gnbrthpl b}
\ee
\be
\dot{p}_{i}=-\frac{\partial H}{\partial \rho_{i}}=\frac{\ell_{i}^{2}}{g_{i}\rho_{i}^{3}}-\frac{\partial V}{\partial \rho_{i}},\label{gnbrthpl c}
\ee
\be
\dot{\ell}_{i}=-\frac{\partial H}{\partial \theta_{i}}=-\frac{\partial V}{\partial \theta_{i}},\label{gnbrthpl d}
\ee
\endnumparts
where $\rho_{i}$, $\theta_{i}$, $p_{i}$ and $\ell_{i}$ are the radius, angle, linear momentum, and angular momentum, respectively, 
of the $i$th reduced mass, for $i=2,\ldots,n$. The potential is defined to be
\be
V=-\sum_{i,j=1\atop{i< j}}^{n} \frac{G m_{i}m_{j}}{r_{ij}} \label{thisisapotentialincaseyouhavenotnoticed}
\ee
and the total kinetic energy is 
\be
K=\frac{1}{2}\sum_{i=2}^{n} \( \frac{p_{i}^{2}}{g_{i}} + \frac{\ell_{i}^{2}}{g_{i}\rho_{i}^2} \). \label{andthisisthekineticenergy}
\ee
It is easy to verify that the Hamiltonian  
$H=K+V$ is conserved by \Eqs{gnbrthpl}.
The total angular momentum $\sum_{i=2}^{n} \ell_{i}$ is also conserved, and the center of mass remains at the origin for all time. 

\subsection{Integration}

Transform $(\v{\rho}, \v{\theta}, \vp,\v{\ell})$ to  $(\v{\zeta},
\v{\theta}, \v{\eta}, \v{\ell})$, where
\numparts
\be\label{gnbtransformedvariables}
\zeta_{2}=V, \label{gnbtransformedvariables a}
\ee
\be
\zeta_{i}=\rho_i, \qquad \hbox{for \  $i=3,\ldots,n$,} \label{gnbtransformedvariables b}
\ee
\be
\eta_{i}=\frac{p_{i}^{2}}{2g_{i}}+\frac{\ell_{i}^{2}}{2g_{i}\rho_{i}^{2}}, \qquad \hbox{for \  $i=2,\ldots,n$.} \label{gnbtransformedvariables c}
\ee
\endnumparts
Note that $H$ is a linear function of the transformed variables:
\be
H=\sum_{i=2}^{n} \eta_{i} + \zeta_{2}, \label{aHamiltonian}
\ee
as is the total angular momentum $L=\sum_{i=2}^{n} \ell_{i}.$
The time derivatives of $\v{\zeta}$ and $\v{\eta}$ are given by
\numparts
\be\label{gnbderivativesoftransformedvariables}
\dot{\zeta}_{2}=\sum_{i=2}^{n}\(\frac{\partial V}{\partial \rho_{i}}\dot{\rho}_i
 + \frac{\partial V}{\partial \theta_{i}}\dot{\theta}_i \),\label{gnbderivativesoftransformedvariables a}
\ee
\be
\dot{\zeta}_{i}=\dot{\rho}_i,  \qquad \hbox{for \  $i=3,\ldots,n$,}\label{gnbderivativesoftransformedvariables b}
\ee 
\be
\dot{\eta}_{i}=\frac{p_{i}\dot{p_{i}}}{g_{i}} + \frac{\ell_{i}\rho_{i}^{2}\dot{\ell}_{i}-\rho_{i}\ell_{i}^{2}\dot{\rho}_{i}}{g_{i}\rho_{i}^{4}}, \qquad \hbox{for \  $i=2,\ldots,n$.}\label{gnbderivativesoftransformedvariables c}
\ee
\endnumparts
The predictor equations are
\numparts
\be\label{gnbpredictors}
{\tilde{\rho}_{i}}=\rho_{i}+\dot{\rho}_{i}\tau,
\qquad
{\tilde{\theta}_{i}}=\theta_{i}+\dot{\theta}_{i}\tau,\label{gnbpredictors a}
\ee
\be
{\tilde{p}_{i}}=p_{i}+\dot{p}_{i}\tau,
\qquad
{\tilde{\ell}_{i}}=\ell_{i}+\dot{\ell}_{i}\tau \label{gnbpredictors b}
\ee
\endnumparts
and the corrector is given by
\numparts
\be\label{gnbcorrectors}
\zeta_{i}(t+\tau)=\zeta_{i}+\frac{\tau}{2}(\dot{\zeta}_{i}+\dot{\tilde{\zeta}}_{i}),
\qquad
\theta_{i}(t+\tau)=\theta_{i}+\frac{\tau}{2}(\dot{\theta}_{i}+\dot{\tilde{\theta}}_{i}),\label{gnbcorrectors a}
\ee
\be
\eta_{i}(t+\tau)=\eta_{i}+\frac{\tau}{2}(\dot{\eta}_{i}+\dot{\tilde{\eta}}_{i}),
\qquad
\ell_{i}(t+\tau)=\ell_{i}+\frac{\tau}{2}(\dot{\ell}_{i}+\dot{\tilde{\ell}}_{i}),\label{gnbcorrectors b}
\ee
\endnumparts
for  $i=2,\ldots,n.$

One then inverts to get the original variables as functions of the temporary transformed variables:
\numparts
\be\label{gnbbacktotheoriginalequations}
\rho_{i}=\zeta_{i} \qquad \hbox{for \  $i=3,\ldots,n$,} \label{gnbbacktotheoriginalequations a}
\ee
\be
\rho_{2}=g(\zeta_{2},\rho_3,\ldots, \rho_n,\v{\theta}), \label{gnbbacktotheoriginalequations b}
\ee
\be
p_{i}=\sgn(\tilde{p_{i}})\sqrt{2g_{i}\(\eta_{i}-\frac{\ell_{i}^{2}}{2g_{i}\rho_{i}^{2}}\)}, \qquad \hbox{for \  $i=2,\ldots,n$.}
\ee
\endnumparts
The value of the inverse function $g$ defined by
\be
V(g(\zeta_{2},\rho_3,\ldots, \rho_n,\v{\theta}),\rho_3,\ldots,
\rho_n,\v{\theta})=\zeta_2
\ee
is determined
at fixed $\rho_3,\ldots,\rho_n$, $\v{\theta}$ with a Newton--Raphson method,
using the predicted value $\tilde {\rho}_2$ as an initial guess.

\begin{figure}
\begin{center}
\includegraphics[height=7.5cm]{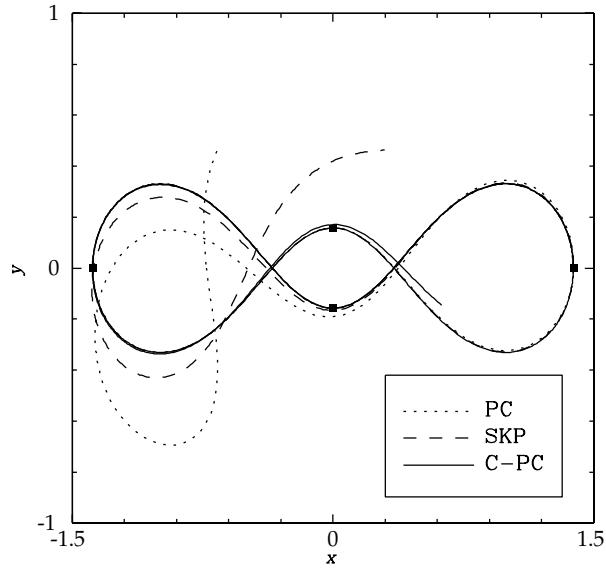}
\caption{The predictor--corrector (dotted line), symplectic (dashed line)
and conservative predictor--corrector (solid line) solutions for a 
four-body choreography.
}
\label{og4b}
\end{center}
\end{figure}

In Fig.~\ref{og4b}, we illustrate Sim\'{o}'s four-body choreography
\cite{Simo00}. The motions of one of
the four unit masses as determined by the predictor--corrector and
conservative predictor--corrector algorithms are compared, using the fixed time step
$\tau=10^{-3}$ to integrate the system from time $t=0$ to $t=11.5$. 
The gravitational constant $G$ is taken to be unity and
the initial conditions are given by
\begin{eqnarray}
\v r_1&=&(1.382857,0),\quad \v r_2=(0,0.157030),\nonumber\\
\v r_3&=&(-1.382857,0),\quad \v r_4=(0,-0.157030),\nonumber\\
\dot{\v r_1}&=&(0,0.584873),\quad \dot{\v r_2}=(1.871935,0),\nonumber\\
\dot{\v r_3}&=&(0,-0.584873),\quad \dot{\v r_4}=(-1.871935,0).
\end{eqnarray}

As $\tau$ is decreased, the predictor--corrector solution converges to the
conservative predictor--corrector solution obtained with a large time
step. This emphasizes that the conservative predictor--corrector can be
viewed as a finite-time-step generalization of the conventional
predictor--corrector, as argued in Ref.~\cite{Shadwick99}. 
We also compare these solutions to a symplectic map
based on the simple second-order kinetic--potential energy splitting 
\begin{eqnarray}
\tilde p_i&=&p_i-\frac{\tau}{2}
\frac{\partial}{\partial q_i} V(q_1,q_2,\ldots, q_N),\nonumber\\
q'_i&=&q_i+\tau\frac{\partial}{\partial \tilde p_i}
K(\tilde p_1,\tilde p_2,\ldots, \tilde p_N),
\nonumber\\
p'_i&=&\tilde p_i-\frac{\tau}{2} V(q'_1,q'_2,\ldots, q'_N),
\end{eqnarray}
to evolve the canonical variables $(q_i,p_i)$ to $(q'_i,p'_i)$, for
$i=1,\ldots,N$. This second-order scheme, which is implemented as the
method SKP using Varadi's {\tt NBI} code with $\tau=10^{-3}$, is similar to
the one described by Forest and Ruth \cite{Ruth83,Forest90}, with the
roles of the coordinates and momenta interchanged.

\begin{figure}
\begin{center}
\includegraphics[height=7.5cm]{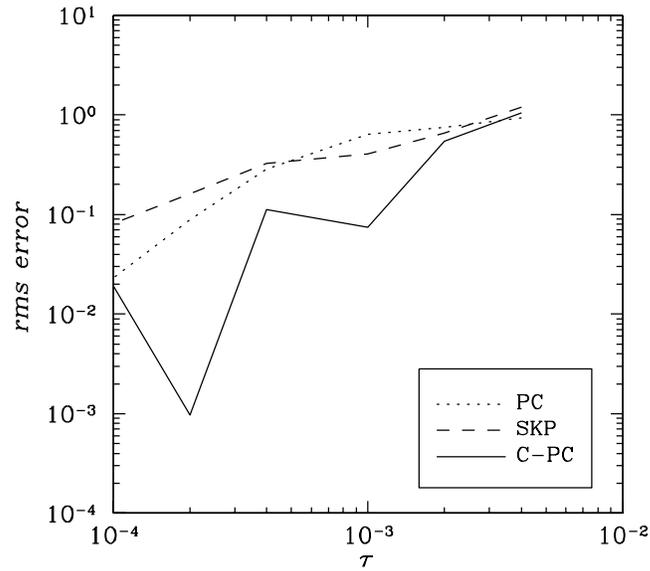}
\caption{Root-mean-square error of the predictor--corrector (dotted line),
symplectic (dashed line) and conservative predictor--corrector
(solid line) solutions in Fig.~\ref{og4b}.}
\label{eg4b}
\end{center}
\end{figure}

In Figure~\ref{eg4b}, we compare the root-mean-square error in the computed
trajectory between $t=0$ and $t=4\pi$ (twice the period of the
choreography), for each of these integration algorithms. The error was
computed relative to a fifth-order Runge--Kutta integrator with time step
$\tau=10^{-5}$. Of the three other solutions, we note that the conservative
predictor--corrector trajectory is the most accurate. For general $n$-body
integrations, our conservative algorithm was also observed to be more
accurate than the second-order symplectic Wisdom--Holman scheme
\cite{Wisdom91,Varadi95,Varadi99}, but this is expected since the latter
applies only to small perturbations of Keplerian orbits.

\section{Conclusion}
Conservative integration algorithms can reduce the computational effort
required to integrate a system of ordinary differential equations.  For
example, when the total energy and angular momentum of the $n$-body problem
is conserved, it is possible to obtain accurate trajectories using a larger
time step than with conventional integration methods.  This is particularly
relevant for extremely long-time integrations.  In contrast, symplectic
methods typically predict a total energy that oscillates about the correct
value. In some cases, these oscillations can eventually lead to large
excursions from the mean value, similar to random walk diffusion.  
In addition, there are certain statistical mechanical systems (such as
equipartition states of inviscid fluids) where the final mean state is a
function of only the initial values of the invariants; for these systems, a
conservative integrator is clearly preferable to a symplectic algorithm.
However, as these methods put the integration error into different places,
the integration method that is most suitable for a given system ultimately
depends on the nature of the physical problem, the integration time
scale, and the kinds of questions addressed by the
numerical simulation.

In the case of the $n$-body problem for planar motion, there are six
invariants, all of which need to be considered during the
integration. Jacobi coordinates were used to reduce the
system to an $(n-1)$-body problem in which the linear momentum and center
of mass constraints are implicitly built in, leaving fewer conservation
laws to be explicitly built into the algorithm. In Jacobi
coordinates, the kinetic energy term of the Hamiltonian remains in diagonal
form (a sum of squares); this makes it easy to express the Hamiltonian as
a linear function of new variables.

Future work in this area should include extending the numerical code to the
full three-dimensional case and regularizing the potential terms to
handle collisions and close approaches. One could also build in precession,
nutation, and tidal effects into the equations of motion.

\bigskip

This work was supported by the Natural Sciences and Engineering Research
Council of Canada.

\appendix
\section{Error Analysis}\label{error}
Here we describe the local error analysis of the second-order
conservative predictor--corrector scheme given by~\Eq{predictorcorrector a}
and~\Eq{cpcb}. We
assume that both $f$ and $T$ are analytic functions and that the points
where $T'$ vanishes are isolated. For notational simplicity, we restrict
the analysis to the autonomous one-dimensional ordinary differential
equation $dx/dt=f(x)$, for which the exact solution is given by
\be
x(t+\t)=x_0+\t f(x_0)+\fr{\t^2}{2}f'(x_0)f(x_0)+\fr{\t^3}{6}f''(x_0)f(x_0)+
\O(\t^4).
\ee
The conservative predictor--corrector scheme
\be
\tilde{x}=x_{0}+\t f(x_0),
\ee
\be
\x(t+\t)=\x_0+\fr{\t}{2} \[T'(x_0) f(x_0)+ T'(\tilde{x}) f(\tilde{x})\]
\ee
yields the solution
\begin{eqnarray}
\x(t+\t)&=&\x_0+\fr{\t}{2} \[T'f+ T'f+\(T'f\)'\t f+
\(T'f\)''\fr{\t^2 f^2}{2}+\O(\t^3) \]\nonumber\\
&=&\x_0+\t T'f+\fr{\t^2}{2}\(T''f^2+T'f'f\)\nonumber\\
&&\quad +\fr{\t^3}{4} \(T'''f^3+2T''f'f^2+T'f''f^2\)+\O(\t^4),
\end{eqnarray}
where the expressions on the right-hand side are all evaluated at $x_0$.
The new value of $x$ is then given by
\begin{eqnarray}
x(t+\t)&=&T\inv\(\x(t+\t)\)
\nonumber\\
&=&T\inv(\x_0)+T\inv{}'(\x_0)\[\t T'f+\fr{\t^2}{2}\(T''f^2+T'f'f\)
\right.\nonumber\\
&&\quad +\left.\fr{\t^3}{4} \(T'''f^3+2T''f'f^2+T'f''f^2\)+\O(\t^4)\]
\nonumber\\
&&\quad +\fr{1}{2} T\inv{}''(\x_0)\[\t T'f+\fr{\t^2}{2}\(T''f^2+T'f'f\)
+\O(\t^3)\]^2
\nonumber\\
&&\quad +\fr{1}{6} T\inv{}'''(\x_0)\[\t T'f+\O(\t^2)\]^3+\O(\t^4).
\label{Taylor}
\end{eqnarray}
By implicitly differentiating the identity $T\inv(T(x))=x$, it follows that
\begin{eqnarray}
T\inv{}'(\x_0)&=&\fr{1}{T'},\quad T\inv{}''(\x_0)=-\fr{T''}{T'^3},
\quad T\inv{}'''(\x_0)=\fr{3T''^2}{T'^5}-\fr{T'''}{T'^4},
\end{eqnarray}
so that \Eq{Taylor} simplifies to
\begin{eqnarray}
x(t+\t)&=&x_0+\t f(x_0)+\fr{\t^2}{2}f'(x_0)f(x_0)
\nonumber\\
&&\quad+\fr{\t^3}{4}\[f''(x_0)f^2(x_0)+\fr{T'''(x_0)}{3T'(x_0)} f^3(x_0)\]
+\O(\t^4).
\label{cpcsoln}
\end{eqnarray}
By setting $T(x)=x$, we obtain the usual error estimate for the conventional
predictor--corrector. We see that both conservative and conventional
schemes are accurate to second order in $\t$; moreover, for quadratic
transformations like $T(x)=x^2$, which in light of Lemma~\ref{linear}
are often useful for enforcing energy conservation, the conservative and 
conventional schemes agree through third order in $\t$. The appearance of
$T'(x_0)$ in the denominator of the third-order (error) term emphasizes that we
must exercise care at singular points of $T$. Near these points, either a
conventional scheme can be used or the time step can be reduced, as previously
remarked \cite{Bowman97}. Should $T'(x_0)=0$, the estimate \Eq{cpcsoln}
should be replaced by
\be
x(t+\t)=T\inv\(T(x_0)+\fr{\t}{2} T'(\tilde{x})f(\tilde{x})\),
\ee
which is guaranteed to have a solution for sufficiently small $\t$ if
the points at which $T'$ vanishes are isolated. The transformation is
then invertible at $x(t+\t)$, allowing the integration to be continued
beyond the point of singularity.

\pagebreak[4]
\centerline{REFERENCES}
\bigskip


\end{document}